\documentstyle[11pt]{article}
\newcommand{\be}{\begin{equation}}
\newcommand{\ee}{\end{equation}}

\newcommand{\bea}{\begin{eqnarray}}
\newcommand{\eea}{\end{eqnarray}}

\begin{document}

\begin{flushright}
IFIC/97--98 
\end{flushright} 

\vspace{0.5cm}

\begin{center}{\Large \bf 
Translationally-invariant coupled-cluster method for finite systems
}\end{center}

\vspace{0.5cm}

\begin{center}{\large 
R. Guardiola, I. Moliner \\ 
{\small Dept.\ de F\'{\i}sica At\'omica, Molecular i Nuclear,
Universitat de Val\`encia} \\ 
{\small Avda.\ Dr.\ Moliner 50, E-46.100 Burjassot, Spain} \\ 
\vspace{0.3cm}
J. Navarro, M. Portesi \\ 
{\small IFIC (Centre Mixt CSIC -- Universitat de Val\`encia)} \\ 
{\small Avda.\ Dr.\ Moliner 50, E-46.100 Burjassot, Spain}
}\end{center}

\vspace{0.2cm}

\begin{abstract}

The translational invariant formulation of the coupled-cluster method is
presented here at the complete SUB(2) level for a system of nucleons
treated as bosons. The correlation amplitudes are solution of a non-linear
coupled system of equations. These equations have been solved for light
and medium systems, considering the central but still semi-realistic
nucleon--nucleon S3 interaction.

\hfill



\end{abstract}

\thispagestyle{empty}

\newpage

\section{Introduction}

The coupled-cluster method (CCM) was invented many years ago by Coester
and K\"ummel \cite{coe58,coe60} to calculate the ground-state energy of
closed-shell nuclei. Since then the CCM has been developed for
applications in a wide variety of fields, including quantum chemistry,
quantum lattices, electron gas, or nuclear physics. Although the formalism
was initially developed to calculate only the energy of the ground state
of a closed-shell nucleus, the CCM is not just a theory for the ground
state of closed-shell systems. Further extensions allow to calculate also
excited states, open-shell systems, or finite temperature systems, to
mention only a few developments of CCM (see general references in Refs.\
\cite{kum78,bis91}). Currently, the CCM provides a widely applicable
formalism to deal with the general quantum many-body problem at a
completely microscopic level. It works for systems of both bosons and
fermions, regardless of the type and range of the interaction.

The CCM wave function is described in terms of the action of a cluster
operator on some reference function, which takes into account the required
symmetry properties of the system under consideration. The cluster
operator is sum of $n$-body operators, which create $n$ particle--hole
pairs on the reference function. In the application to finite systems, one
faces with the well-known center-of-mass motion problem. The proper
treatment of the center-of-mass in the framework of CCM was initiated some
years ago \cite{bis90,bis92} at the so-called SUB(2) level of
approximation, which is limited up to two-body operators. It has been
shown in these references that the center-of-mass is properly removed
using appropriate combinations of one- and two-body operators, and
describing the reference function in terms of single-particle harmonic
oscillator wave functions. Such reformulation of the CCM is called
translationally invariant coupled-cluster (TICC) method, and it has been
applied to the $^4$He nucleus, considered as a system of four bosons. For
the simple interactions considered in Ref.\ \cite{bis92}, it was shown
that the results at the SUB(2) level (TICC2 henceforth) may differ in a
few MeV from the essentially exact diffusion Monte Carlo results.

The CCM is most naturally formulated in the occupation number
representation. However, very slow convergence with respect to the cut-off
on the single particle basis has been found. This led to consider a
linearized form of the TICC2 approach, which can be easily converted to
coordinate representation. The pair correlations are considered through a
function depending on the relative coordinate of a single pair. The ansatz
for the wave function of a fermionic system is 
$$ 
|\Psi\rangle =
\sum_{i<j} \left( f_c(r_{ij}) + f_{\sigma}(r_{ij}) (\sigma_i \cdot
\sigma_j) + f_{\tau}(r_{ij}) (\tau_i \cdot \tau_j) + f_{\sigma
\tau}(r_{ij}) (\sigma_i \cdot \sigma_j)(\tau_i \cdot \tau_j)
\right) |\Phi\rangle, 
$$
where $|\Phi\rangle$ is the reference state. The four unknown two particle
correlation functions may be determined by minimizing the expectation
value of the hamiltonian, and a single correlation function appears in the
case of a system of bosons. This so-called translationally invariant
configuration interaction method was applied in Ref.\ \cite{lineal} to
calculate at the SUB(2) level (or TICI2 approach) the ground-state energy
of some light nuclei in the $1p$-shell up to $^{16}$O. The main conclusion
of this work is that the TICI2 methodology provides a very reasonable
starting point for the calculation of the binding energies of
light-to-medium nuclei. For interactions and correlations of the V4 form,
as displayed in the above equation, the TICI2 results are in suitable
agreement with the ones provided by other methodologies.

In order to get a better description of finite nuclei there are however
some obvious extensions to be performed. Among them are the inclusion of
all the correlations at the SUB(2) level, i.e.\ the TICC2 approximation.
In Refs.\ \cite{bis90,bis92} it has been shown that the additional terms
give a relatively small contribution to the ground-state energy of the
$^4$He nucleus. The purpose of the present work is to discern the
importance of the non-linear terms beyond TICI2 in the description of a
bosonic system. In Section 2 we briefly summarize the main findings of
previous works \cite{bis90,bis92}, generalizing at the same time the
results for $N$ bosons. In Section 3 the detailed equations are presented
and discussed. An application example is given in Section 4, where a
simple nuclear interaction is used to determine the ground-state energy of
a system of nucleons treated as bosons. Finally, in Section 5 some
concluding remarks and an outlook for further studies are given.

\section{The Translationally Invariant SUB(2) approximation for bosons}

The basic CCM ansatz is to write the $N$-body wave function as
\be
\vert \Psi \rangle = {\rm e}^S \vert \Phi \rangle, 
\label{ccmwf}
\ee
where $\vert \Phi \rangle$ is a reference state, incorporating the
statistical and symmetry properties of the system, and
\be
S = \sum_{n=1}^N S_n
\label{correl}
\ee
is the cluster correlation operator, sum of operators of the form
\be
S_n = \frac{1}{(n!)^2} \sum_{\rho_1,\ldots ,\rho_n} \langle \rho_1,\ldots,
\rho_n | S_n | 0,\ldots,0 \rangle a^+_{\rho_1} \ldots a^+_{\rho_n} a^n_{0},
\label{corren}
\ee
whose action is to create $n$ particle--hole excitations on the reference
function. The SUB(2) level approximation of CCM consists of neglecting the
excitations of more than two particle--hole pairs, so that in the initial
ansatz one has $S=S_1+S_2$.

In order to get rid of the center-of-mass problem, one could directly
choose a translationally invariant reference state, but this will spoil
the meaning of the Fock space operator as single-particle operators, due
to the center-of-mass constraint. The alternative followed in Ref.\
\cite{bis90} was to build up the reference state from harmonic-oscillator
(HO) single-particle wave functions, which can be defined through creation
operators acting on the vacuum,
\be
|nlm\rangle = a^+_{nlm} |0\rangle, 
\label{spho}
\ee
and to write the reference state of an $N$-boson system as
\be
|\Phi \rangle = \frac{1}{\displaystyle \sqrt{N!}} (a^+_{000})^N
|0\rangle .
\label{howf}
\ee
It is well-known that this HO uncorrelated wave function is the simple
product of a $1s$ HO wave function for the center-of-mass and an intrinsic
wave function. Although the reference state is not translationally
invariant, this factorization property allows one to remove the
center-of-mass motion unambiguously. Dressing the reference state with
correlations has the danger of spoiling the factorization property, unless
the correlations are translationally invariant. Therefore our immediate
goal is to construct excitations on $|\Phi\rangle$ without exciting the
center-of-mass, so that in this way it can be properly removed from the
SUB(2) wave function.

With only $S_1$ cluster operators it is not possible to get
translationally invariant correlations. However, appropriate admixtures of
$S_1$ and $S_2$ can do it. Let us indicate by $S^{(1,2)}$ the
corresponding new cluster operator. Translational invariance may be
imposed by: i) recoupling the product of single-particle
harmonic-oscillator states into sums of comparable products of oscillator
states for the relative and center-of-mass motion of the pair, and ii)
imposing that the center-of-mass state of the destroyed pair of particles
in the occupied subspace is the same as that of the created pair of
particles in the unoccupied subspace. Since we are using HO wave
functions, this is accomplished by using the Brody--Moshinsky brackets
\cite{bro67}. In the case of an $N$ boson system, one can write the
cluster operator as
\be
S^{(1,2)} = \sum_{n=1}^{\infty} {\cal S}(n) \sum_{n_1, n_2, l} 
\langle n0,00,0|n_1l, n_2l, 0\rangle \left[ a^+_{n_1l} \times
a^+_{n_2l} \right]^0_0 a^2_{000},
\label{ese12}
\ee
where $\langle n0,00,0|n_1l, n_2l, 0\rangle$ is a Brody--Moshinsky
coefficient, which limits the sums to the cases $n_1+n_2+l=n$. This
expression reflects that two bosons in occupied states of $|\Phi\rangle$
are destroyed, and the created pair of particles has zero angular
momentum, its center-of-mass being in the HO state with all quantum
numbers equal to zero. Note that the term with both $(n_1,l)=(0,0)$ and
$(n_2,l)=(0,0)$ is excluded, as it simply reproduces the uncorrelated
state $|\Phi\rangle$. However, the terms with either $(n_1,l)=(0,0),
n_2\neq 0$ or $n_1\neq 0, (n_2,l)=(0,0)$ are included. These terms give
precisely the 1p--1h excitations in the admixture.

It is worth noting the enormous simplification imposed by the
translational invariance, because we have a single correlation amplitude
${\cal S}(n)$, whose argument is just a single number which counts the
number of oscillator quanta (divided by 2) globally excited. By contrast,
in the unrestricted SUB(2) approximation, one deals with two correlation
amplitudes ${\cal S}_1(n)$ and ${\cal S}_2(n_1,n_2,l)$.

It is convenient to use the following simplified notation for the
correlation operator:
\be
S^{(1,2)} = s(p) a^+_p a^+_0 a^2_0 + s(p, q) a^+_p a^+_q a^2_0,
\label{s12simp}
\ee
where a sum over the particle indices $p, q$ is to be assumed. These
indices refer to the HO quantum numbers: $p\equiv(n_p,l_p,m_p)$, and they
are different from (0,0,0), since they correspond to non-occupied states.
For convenience, we shall also use the notation $0\equiv (0,0,0)$. The
coefficients are defined as
\be
s(p) = \delta(l_p,0) \delta(m_p,0) 2 
\langle n_p0,00,0|n_p0,00,0\rangle {\cal S}(n_p),
\label{esep}
\ee
\bea
s(p, q) &=& \delta(l_p,l_q) \delta(m_p,-m_q)
C(l_p,l_p,0;m_p,-m_p,0) \nonumber \\ 
&& \langle n_p+n_q+l_p0,00,0|n_pl_p,n_ql_p,0\rangle {\cal S}(n_p+n_q+l_p),
\label{esepq}
\eea
where $C(l_1,l_2,l_3;m_1,m_2,m_3)$ is a Clebsh--Gordan coefficient.

Finally, let us note that simply because $S^{(1,2)}$ is invariant under
translations we are by no means guaranteed that so does the exponential
form ${\rm e}^{S^{(1,2)}}$. This can be easily seen by considering the
simple example of the squared correlation operator $S^{(1,2)} S^{(1,2)}$.
Indeed, one of the destruction operators $a_0$ of one factor may be
contracted with the creation operator $a^+_0$ of the other factor,
resulting in a term which is not translationally invariant. As was shown
in Ref.\ \cite{bis90} the unwanted terms are easily excluded by the simple
device of taking the ${\rm e}^{S^{(1,2)}}$ operator in normal-ordered
form, so that the TICC2 ansatz for the wave function can be finally
written as
\be
\vert \Psi \rangle = \, :{\rm e}^{S^{(1,2)}}: \vert \Phi \rangle .
\label{ticc2}
\ee
In principle, for a system of bosons the normal ordered form does not
imply a particular technical problem (contrary to the fermionic case).
However, the usual CCM equations are more involved, and we found it more
convenient to derive the CCM equations in a different manner, as it will
be shown in the next section.

\section{The TICC2 equations}

The usual way to solve the Schr\"odinger equation
\be
H |\Psi\rangle = E |\Psi\rangle
\label{schro}
\ee
within CCM involves transforming it first to
\be
{\rm e}^{-S} H {\rm e}^{S} |\Phi\rangle = E |\Phi\rangle,
\label{inver}
\ee
i.e.\ to define a similarity transformed hamiltonian. The motivation to
use this equation is that the transformed hamiltonian ${\rm e}^{-S} H {\rm
e}^{S}$ may be reduced to a set of nested commutators, which may result in
a simplification of the algebra. Afterwards, Eq.\ (\ref{inver}) is
projected onto 0p--0h, 1p--1h, $\ldots$ states to get algebraic equations
for the amplitudes. If we were projecting up to $N$p--$N$h, the
corresponding set of algebraic equations would be equivalent to the
Schr\"odinger equation. However, if the cluster operator is truncated at a
given SUB($n$) level the projection up to $N$p--$N$h states will produce
redundant equations, so the current practice is to project up to the
$n$p--$n$h states. It is worth noting that if the cluster operator is
truncated at the SUB($n$) level none of Eqs.\ (\ref{schro}) or
(\ref{inver}) may be exactly satisfied, and moreover one should not expect
in general to obtain the same results from both equations.

The similarity transformation does not appear convenient in our formalism
with the normal ordering prescription on $\exp(S)$. Certainly, the inverse
of the operator $:\exp(S):$ is not $:\exp(-S):$, but it is possible to
find a new operator $T$ such that the equation
\be
T \, :{\rm e}^{S}: |\Phi \rangle = |\Phi \rangle
\ee
is satisfied. For the specific operator $S^{(1,2)}$ given in Eq.\
(\ref{s12simp}), there results
\be
T \simeq 1 - (N-1) s(p) a^+_p a_0 + \left[ - s(p,q) +
\frac{1}{2}(N+1)(N-2) s(p) s(q) \right] a^+_p a^+_q a^2_0 + \ldots,
\label{late}
\ee
up to two creation operators. This operator $T$ is not the inverse of
$:\exp(S^{(1,2)}):$ because it is defined when acting on the reference
state $|\Phi\rangle$, and this fact is reflected in its dependence on the
number of bosons of the system. The structure of $T$ depends on the
reference state, because also normal ordering depends on such state.

Another much simpler alternative to obtain the equations for the
amplitudes is to project directly the Schr\"odinger equation with the
ansatz (\ref{ticc2})
\be
H \, :{\rm e}^{S^{(1,2)}}: | \Phi \rangle = E \, :{\rm e}^{S^{(1,2)}}:
| \Phi \rangle
\label{schro2}
\ee
onto the states with 0p--0h, 1p--1h and 2p--2h. 

For hamiltonians with two-body interaction potentials like
\be
H = \sum_{\alpha,\beta} \langle \alpha | K | \beta \rangle
a^+_{\alpha} a_{\beta} + \frac{1}{4}
\sum_{\alpha_1,\alpha_2,\beta_1,\beta_2} \langle \alpha_1,
\alpha_2 | V | \beta_1, \beta_2 \rangle a^+_{\alpha_1}
a^+_{\alpha_2} a_{\beta_2} a_{\beta_1}
\label{hamil}
\ee
one requires the expansion of $:\exp(S^{(1,2)}):$ up to terms involving at
most four particle creation operators. In terms of the quantities $s(p)$
and $s(p,q)$ defined in Eqs.\ (\ref{esep}) and (\ref{esepq}),
respectively, one may effectively do the substitution
\bea
:{\rm e}^{S^{(1,2)}}: &=& 1 + s(p)a^+_p a^+_0 a^2_0 + s(p,q) a^+_p
a^+_q a^2_0 +\frac{1}{2} s(p) s(q) a^+_p a^+_q (a^+_0)^2 a_0^4
\nonumber \\ 
&&+ s(p) s(q,r) a^+_p a^+_q a^+_r a^+_0 a_0^4 
+ \frac{1}{6} s(p) s(q) s(r) a^+_p a^+_q a^+_r (a^+_0)^3 a_0^6
\nonumber \\
&&+ \frac{1}{2} s(p,q) s(r,s) a^+_p a^+_q a^+_r a^+_s a_0^4 
+\frac{1}{2} s(p) s(q) s(r,s) a^+_p a^+_q a^+_r a^+_s (a^+_0)^2 a_0^6
\nonumber \\
&&+ \frac{1}{24} s(p) s(q) s(r,s) a^+_p a^+_q a^+_r a^+_s (a^+_0)^4 a_0^8
+ \ldots,
\eea
which is no more complex than the usual nested commutators. Note the
reorderings of $a_0$ and $a^+_0$ related to the normal ordering
prescription.

The projection onto 0p--0h is simply the projection onto the bra state
$\langle \Phi |$ defined by the reference state, and it gives the
ground-state energy:
\be
E = \langle \Phi | H \, :{\rm e}^{S^{(1,2)}}: | \Phi \rangle .
\label{gse}
\ee
Using the following notation:
\bea
k(\alpha,\beta) &=& \langle n_{\alpha} l_{\alpha} m_{\alpha} | K
| n_{\beta} l_{\beta} m_{\beta} \rangle, 
\label{kine} \\
v(\alpha_1,\alpha_2,\beta_1,\beta_2) &=& \langle n_{\alpha_1}
l_{\alpha_1} m_{\alpha_1} ; n_{\alpha_2} l_{\alpha_2}
m_{\alpha_2} | V | 
n_{\beta_1} l_{\beta_1} m_{\beta_1} ; n_{\beta_2} l_{\beta_2}
m_{\beta_2} \rangle,
\label{pote}
\eea
for the matrix elements of kinetic and potential energies taken between HO
single particle wave functions, we find
\bea
E &=& N_1 k(0,0) + N_2 v(0,0,0,0) \nonumber \\
&+& N_2 \bigg[ k(0,p) s(p) + 2 (N-1) v(0,0,p,0) s(p) + 2 v(0,0,p,q)
s(p,q) \bigg] \nonumber \\
&+& N_4 v(0,0,p,q) s(p) s(q),
\label{energia}
\eea
where we have also defined 
\be
N_i = N(N-1) \ldots (N-i+1).
\label{factor}
\ee
Again, a sum over repeated particle indices $p$, $q$ is to be understood
in Eq.\ (\ref{energia}).

The ground-state energy is thus given in terms of the correlation
amplitudes ${\cal S}(n)$ implicit in $s(p)$ and $s(p,q)$. To determine
these amplitudes, we must project the Schr\"odinger equation onto 1p--1h
and 2p--2h states taking care that the center-of-mass state is the same as
in the reference function. The obvious state to project on is the
appropriate 1p--1h and 2p--2h admixture which maintains the
factorizability of the center-of-mass in the reference state. Such an
admixture is already known, after the discussion of Section 2: it must
have the same structure as $S^{(1,2)}$, but without the unknown amplitudes
${\cal S}(n)$. It seems convenient to classify the excitations in terms of
the number of oscillator quanta $2N_x$, where $N_x$ will run from 1 up to
some sufficiently high value $N_{\rm max}$ so as to reach convergence. In
this way one obtains $N_{\rm max}$ non-linear equations to determine the
unknown amplitudes ${\cal S}(n)$. Once these amplitudes are known, the
energy is easily computed.

Consequently, the excited bra state required for the proper projection can
be written as
\be
\langle \Phi | \left[ c(m) (a^+_0)^2 a_0 a_m + 
c(m, n) (a^+_0)^2 a_m a_n \right],
\label{bra}
\ee
where the excitation amplitudes have a structure similar to the cluster
operator, that is
\be
c(m)=\delta(l_m,0) \delta(m_m,0) 2 \langle N_x0,00,0 | n_m0,00,0 \rangle,
\label{cm}
\ee
\be
c(m, n) = \delta(l_m,l_n) \delta(m_m,-m_n) C(l_m,l_m,0;m_m, -m_m,0) 
\langle N_x0,00,0 | n_ml_m, n_nl_m,0 \rangle .
\label{cmn}
\ee
The indices $m$, $n$ refer to excited states, and cannot be identical with
(0,0,0). Moreover, they are external indices, tied to the excitation
quantum number $N_x$, so that there is no summation over them.

Thus, the equation to determine the correlation amplitudes can be written
as:
\bea
&& \langle \Phi | \left[ c(m) (a^+_0)^2 a_0 a_m +
c(m, n) (a^+_0)^2 a_m a_n \right] H
\, :{\rm e}^{S^{(1,2)}}: |\Phi\rangle \nonumber \\
&& \hspace{1truecm}= E \langle \Phi | \left[ c(m) (a^+_0)^2 a_0 a_m +
c(m, n) (a^+_0)^2 a_m a_n \right] 
:{\rm e}^{S^{(1,2)}}: |\Phi \rangle .
\label{project}
\eea
Evaluating all contractions contained in this equation is not a simple
task, because of the very large number of terms which appear after the
expansion of the operators. We have found of great help the use of the
algebraic language REDUCE\footnote{Copyright \copyright 1993 The RAND
Corporation, Santa Monica, CA, USA.}. The core of the evaluation is the
substitution
\be
a_x a^+_y \to a^+_y a_x + \delta(x,y),
\ee
which is controlled by a {\tt let} rule. This automatically reorders the
full expression with all annihilation operators on the right. Afterwards
the Kronecker delta functions are conveniently eliminated.

The left-hand side of Eq.\ (\ref{project}) can be written in the form
\bea
&& \langle \Phi | \left[ c(m) (a^+_0)^2 a_0 a_m +
c(m, n) (a^+_0)^2 a_m a_n \right] H
\, :{\rm e}^{S^{(1,2)}}: |\Phi\rangle \nonumber \\
&=& N_2 c(m) k(m,0) \nonumber \\
&+& N_2 \bigg[
2 (N-2) c(m,n) k(0,0) s(m,n)+ (N-1)^2 c(m) k(0,0) s(m) 
\nonumber \\
&& +2 (N-1) c(m,n) k(m,0) s(n)+2 (N-1) c(m) k(0,p) s(m,p)
\nonumber \\
&& +4 c(m,n) k(m,p) s(n,p)+(N-1) c(m) k(m,p) s(p) \bigg]
\nonumber \\ 
&+& N_4 \bigg[
(N-2) c(m,n) k(0,0) s(m) s(n)+2 c(m,n) k(0,p) s(m,n) s(p)
\nonumber \\ 
&& +4 c(m,n) k(0,p) s(m,p) s(n)+(N-1) c(m) k(0,p) s(m) s(p)
\nonumber \\ 
&& +2 c(m,n) k(m,p) s(n) s(p) \bigg] \nonumber \\
&+& N_6 \bigg[ c(m,n) k(0,p) s(m) s(n) s(p) \bigg] \nonumber \\
&+& N_2 \bigg[ 2 (N-1) c(m) v(m,0,0,0)+2 c(m,n) v(m,n,0,0)
\bigg] \nonumber \\ 
&+& N_2 \bigg[ (N-1)^2 (N-2) c(m) v(0,0,0,0) s(m) \nonumber \\
&& +2 (N-2) (N-3) c(m,n) v(0,0,0,0) s(m,n) \nonumber \\
&& +4 (N-1) (N-2) c(m,n) v(m,0,0,0) s(n) \nonumber \\
&& +4 (N-1) (N-2) c(m) v(0,0,p,0) s(m,p)
\nonumber \\
&&+4 (N-1)^2 c(m) v(m,0,p,0) s(p)
+16 (N-2) c(m,n) v(n,0,p,0) s(m,p)
\nonumber \\
&&+4 (N-1) c(m) v(m,0,p,q) s(p,q)
+4 (N-1) c(m,n) v(m,n,p,0) s(p)
\nonumber \\
&& +4 c(m,n) v(m,n,p,q) s(p,q) \bigg] \nonumber \\
&+& N_4 \bigg[
(N-2) (N-3) c(m,n) v(0,0,0,0) s(m) s(n) \nonumber \\
&& +4 (N-3) c(m,n) v(0,0,p,0) s(m,n) s(p)\nonumber \\
&&+8 (N-3) c(m,n) v(0,0,p,0) s(m,p) s(n) \nonumber \\
&&+2 (N-1) (N-2) c(m) v(0,0,p,0) s(m) s(p) \nonumber \\
&&+4 c(m,n) v(0,0,p,q) s(m,n) s(p,q)
+8 c(m,n) v(0,0,p,q) s(m,p) s(n,q)\nonumber \\
&&+8 (N-2) c(m,n) v(n,0,p,0) s(m) s(p)
+4 (N-1) c(m) v(0,0,p,q) s(m,q) s(p)\nonumber \\
&&+2 (N-1) c(m) v(0,0,p,q) s(m) s(p,q)
+2 (N-1) c(m) v(m,0,p,q) s(p) s(q) \nonumber \\
&&+16 c(m,n) v(n,0,p,q) s(m,q) s(p)
+8 c(m,n) v(n,0,p,q) s(m) s(p,q) \nonumber \\
&&+2 c(m,n) v(m,n,p,q) s(p) s(q) \bigg] \nonumber \\
&+& N_6 \bigg[ 2 (N-3) c(m,n) v(0,0,p,0) s(m) s(n) s(p) \nonumber \\
&&+(N-1) c(m) v(0,0,p,q) s(m) s(p) s(q) \nonumber \\
&&+8 c(m,n) v(0,0,p,q) s(m,p) s(n) s(q)
+2 c(m,n) v(0,0,p,q) s(m,n) s(p) s(q) \nonumber \\
&&+2 c(m,n) v(0,0,p,q) s(m) s(n) s(p,q)
+4 c(m,n) v(n,0,p,q) s(m) s(p) s(q) \bigg] \nonumber \\
&+& N_8 \bigg[ c(m,n) v(0,0,p,q) s(m) s(n) s(p) s(q) \bigg] .
\label{hproject}
\eea
In the right-hand side of Eq.\ (\ref{project}) one has
\bea
&&\langle \Phi | \left[ c(m) (a^+_0)^2 a_0 a_m +
c(m, n) (a^+_0)^2 a_m a_n \right] 
:{\rm e}^{S^{(1,2)}}: |\Phi \rangle \nonumber \\
&=& N_2 \bigg[ (N-1) c(m) s(m) + 2 c(m,n) s(m,n) \bigg] \nonumber \\
&+& N_4 \bigg[ c(m,n) s(m) s(n) \bigg] .
\label{nproject}
\eea
Let us recall that in these equations any index $m$, $n$, $p$, or $q$ must
be interpreted as a set of three quantum numbers $n\equiv(n_n,l_n,m_n)$ of
the HO states, none of them being (0,0,0). Let us insist again that in
these expressions, indices $p,q$ refer to particle states and a sum on
them must be understood, indices $m,n$ are external, and are related to
the quantum number $N_x$ characterizing the excitation.

We have stressed the factors $N_2$, $N_4$, $N_6$, and $N_8$ accompanying
the correlation of one, two, three, and four particle--hole pairs. The
results of Ref.\ \cite{bis90} showed that in the case of $^4$He the
quadratic terms give a relatively unimportant contribution to the
ground-state energy, but we cannot assume this to be the case for larger
systems. Firstly, it is clear that only two pairs can be created for
$N=4$ particles, and consequently the TICC2 approximation for $^4$He
contains only linear and quadratic terms. But more pairs can be excited
for $N>4$, and we see that up to four pairs are present when the
interaction is a two-body one. Secondly, the expressions given above for
the projection show that these terms are multiplied by increasing powers
of $N$: quadratic terms are multiplied by a factor $N_4$, cubic terms by
$N_6$, and quartic terms by $N_8$. Although the amplitudes ${\cal S}(n)$
may depend on $N$ one cannot discard the possibility that the non-linear
terms could become relatively more important for high values of $N$.

We have to solve a non-linear (actually quartic) system of coupled
equations to determine the amplitudes ${\cal S}(n)$, which can be cast in
the following schematic form:
\be
\sum_{i_1,i_2,i_3,i_4=0}^{N_{\rm max}} F(N_x, i_1, i_2, i_3, i_4) 
{\cal S}(i_1) {\cal S}(i_2) {\cal S}(i_3) {\cal S}(i_4) = 0,
\label{cc2eq}
\ee
with the convention ${\cal S}(0)=1$. The number $N_x=1,\ldots,N_{\rm
max}$ refers to the excitation considered. The function
$F(N_x,i_1,i_2,i_3,i_4)$ is obtained from the expressions given above for
the 
projections onto the excitations up to $2N_x$ quanta. The indices $i$ in
the correlation amplitudes refer to the number $2i$ of HO quanta
characterizing the excitation, which we limit to a maximum number $N_{\rm
max}$. The independent term is given by $F(N_x,0,0,0,0)$. The linear
version of the present development, or TICI2 approximation, amounts to
consider in Eq.\ (\ref{cc2eq}) only the function $F(N_x,i_1,0,0,0)$
(including $i_1=0$), and the problem is reduced to solve a generalized
eigenvalue equation.

There is still a long way from Eqs.\ (\ref{hproject}) -- (\ref{nproject})
to the practical equations. Rotational invariance of the hamiltonian
merged with the specific form of $s(p)$ and $s(p,q)$ results in a drastic
reduction on the number of summed up indices. Additionally, the potential
matrix elements are conveniently transformed into Talmi integrals,
referring only to the relative coordinate of the pair. The final formula
is too long to be included here. A part of the non-linear equations, up to
quadratic terms in the amplitudes, may be found in Appendix A of Ref.\
\cite{bis90}, for the $N=4$ case.

\section{An application example}

We consider a model system consisting of $N$ nucleons treated as bosons
interacting via the Wigner part of a nuclear interaction. These systems
collapse for large values of $N$, and so one could reasonably expect that
the effects related to the correlations will be enhanced. We have used the
Wigner part of the Afnan--Tang nucleon--nucleon potential S3 \cite{afnan},
which is considered as a semi-realistic interaction having a quite strong
core at the origin. This will enhance again the role of the correlations.

The price to be paid when going from the configuration interaction TICI2
scheme to the full TICC2 one is that instead of solving a generalized
eigenvalue problem as was done in Ref.\ \cite{lineal}, one has to solve
the non-linear system of equations (\ref{cc2eq}) coupled to the equation
determining the energy $E=\langle \Phi | H \, :\exp(S^{(1,2)}): | \Phi
\rangle$. The obvious steps are first to obtain explicitly the matrix of
coefficients $F(N_x,i_1,i_2,i_3,i_4)$ and afterwards to solve for the
amplitudes and the energy. The equations depend on the harmonic oscillator
parameter $\alpha=(m\omega/\hbar)^{1/2}$ used to define the Fock space, so
it is convenient to adjust this $\hbar \omega$ parameter like in a
variational calculation.

As we have already mentioned, we may limit the number of equations by
means of the index $N_{\rm max}$, which measures the number of excitation
quanta $(2 N_{\rm max} \hbar \omega)$ related to a given term of the
correlation. One may then proceed sequentially, starting from a small
value of $N_{\rm max}$, and taking its solution as the approximate
solution corresponding to the next set of $N_{\rm max}+1$ equations. The
Newton--Raphson method turns out to be very efficient once we are quite
close to the solution.

An alternative way is to introduce in Eq.\ (\ref{cc2eq}) a quenching
factor $q$ as follows:
\bea
&&\sum_{i_1=0}^{N_{\rm max}} F(N_x,i_1,0,0,0) {\cal S}(i_1) \nonumber \\
&+& q \sum_{i_1=0}^{N_{\rm max}} \sum_{(i_2,i_3,i_4)
\neq(0,0,0)}^{N_{\rm max}} F(N_x,i_1,i_2,i_3,i_4) 
{\cal S}(i_1) {\cal S}(i_2) {\cal S}(i_3) {\cal S}(i_4) = 0,
\label{quenching}
\eea
where in the second sum the indices $(i_2,i_3,i_4)$ cannot be
simultaneously equal to zero, as this case is included in the first sum.
We let the quenching factor $q$ slowly grow from 0 to 1, in such a manner
that $q=0$ corresponds to the linear problem and $q=1$ corresponds to the
full quartic set of equations. Solutions of the full equation are thus
smoothly connected to solutions of the simpler linear problem.

There is no a priori way of knowing if all solutions have been determined,
as well as which of these solutions are physically relevant. In the TICI2
formulation there are as many solutions as the number of equations; the
lowest one corresponds to the ground state of the system and the rest
could be interpreted as monopole excitations or breathing modes.

We have solved the non-linear system of equations for $N=$4, 16 and 40
bosons using the S3 interaction. In Fig.\ 1 are displayed the energies of
two selected TICC2 solutions (only one for $N=$4) and their evolution with
the sequential scheme. These two solutions are represented by the solid
and the long-dashed lines. The linear TICI2 solution is also displayed as
the short-dashed line. Several comments are in order regarding these three
solutions.

(i) All of them tend to stability with increasing values of $N_{\rm max}$,
and at $N_{\rm max}=30$ they have already converged.

(ii) There is a connection between the solid line and the TICI2 line in
the quenching scheme and in both directions. This means that starting from
TICI2 solution ($q=0$) and slowly increasing the value of $q$ one arrives
to the solid line TICC2 solution when $q=1$, and conversely, starting from
the solid line TICC2 solution ($q=1$) one arrives to the TICI2 solution by
slowly decreasing $q$ down to $q=0$.

(iii) There is {\em not} a connection path which starts from the
long-dashed TICC2 solution. When $q$ is slowly decreased the solution
disappears.

In conclusion, even if the TICC2 solution represented by the long-dashed
line has the lowest energy value, it should be considered as a spurious
solution of the TICC2 method which has no relation with the physical
problem. This discussion has been quite simplified because we have focused
only on two cases. Actually, by generating at random starting points for
the Newton--Raphson method, there appears a very large number of solutions
which must be discriminated.

The arguments of correctness of solutions, both to increasing $N_{\rm
max}$ or along the quenched path, do not really ascertain the goodness of
a given solution. A precise but costly way is to compute the expectation
value of the hamiltonian for any of the found solutions, but this is a
hopeless task in the Fock representation when a large number of particles
is involved. For this calculation one requires the matrix element
\be
\langle H \rangle = \frac{\langle \Phi | \left( :{\rm e}^{S^{(1,2)}}:
\right)^+ H :{\rm e}^{S^{(1,2)}}: | \Phi \rangle}
{\langle \Phi | \left( :{\rm e}^{S^{(1,2)}}:
\right)^+ :{\rm e}^{S^{(1,2)}}: | \Phi \rangle},
\ee
with a number of contractions scaling as $N^2$.

An alternative way is to go to coordinate representation, where
\bea
\langle {\bf r}_1,\ldots ,{\bf r}_N | :{\rm e}^{S^{(1,2)}}: |
\Phi\rangle &=& \{ 1 + \sum_{i<j} f_{ij} + \frac{1}{2!} \sum_{i<j}
\sum_{k<l}{'} f_{ij} f_{kl} \nonumber \\
&+& \frac{1}{3!} \sum_{i<j}\sum_{k<l}{'}\sum_{m<n}{''} 
f_{ij} f_{kl} f_{mn}+\ldots\}\Phi_{\rm HO}({\bf r}_1,\ldots,{\bf r}_N)
\label{coord}
\eea
In this expression, the correlation function $f$ is defined as
\be
f(r) = \sum_{n=1}^{\infty} {\cal S}(n) \, 2 \left[ \frac{2^n
n!}{(2n+1)!!} \right]^{1/2} L_n^{1/2} ({\textstyle\frac 12} \alpha^2 r^2),
\label{laefe}
\ee
and the sums over single pairs, double pairs, $\ldots$, reflect the
combinatorics of the selection of annihilation operators for the
contraction of the creation operators contained in $:{\exp}(S^{(1,2)}):$.
The primes mean that no repeated particle indices appear in the multiple
sums. The dots in Eq.\ (\ref{coord}) indicate up to $N/2$ sums.

In Fig.\ 2 are displayed the pair correlation functions $f(r)$ for the
solutions considered in Fig.\ 1. With the corresponding wave function
given in Eq.\ (\ref{coord}) we have computed the expectation value of the
Hamiltonian by means of the Monte Carlo method with a Metropolis sampling,
and the reader should be aware that the sampling with a wave function like
(\ref{coord}) is very costly in terms of computation time.

The simplest (and test) case corresponds to $N=4$. There we have not found
spurious solutions, and the variational Monte Carlo (VMC) estimate for the
energy corresponding to the lowest-energy TICC2 solution is $-25.55\pm
0.05$ MeV, in a very good agreement with the normal TICC2 value, which is
displayed in Table 1.

The test has also been carried out for the normal and the spurious
solutions of $N=16$. For the solid line solution the VMC value is
$-1207\pm 16$ MeV, in good agreement with the normal TICC2 value, which is
displayed in Table 1. However, the VMC value for the long-dashed solution
turns out to be $-661\pm 171$ MeV, with a very large variance, and very
different from the corresponding TICC2 value of $-1638$ MeV. This is, in
our mind, a definitive argument to consider the long-dashed line of Figs.\
1--2 as a spurious solution, which should be ruled out. It is very
interesting the agreement between TICC2 and TICC2--VMC calculations, and
it is particularly stimulating if we realize that the TICC2 method is {\em
not} variational. The closeness of both results means that TICC2
determines appropriately the energy.

The non-variational character of the SUB($n$) truncations of CCM should
not be considered as a drawback. Variational methods are very popular in
many-body calculations, but the only advantage of using them, apart from
obtaining an upper bound, is just to secure an improvement of the
ground-state energy when the Hilbert space is enlarged. On the other hand,
the hierarchical structure of the SUB($n$) truncation has some
similarities with perturbation theory, in the sense that the exact value
is not monotonically approached. Both methods may be quite far away from
the exact (and unknown) ground-state energy: in Jastrow case because of
using an inadequate correlation, and in CCM because of a too early
truncation.

It is worth mentioning that the problem of spurious solutions of the CCM
method is one of the open questions related to this theory. In some simple
cases \cite{mark}, it has been possible to study the full map of
solutions, but nevertheless these studies do not help too much in
understanding the real origin of the spurious solutions. Here, the claimed
advantage of the CCM, i.e.\ its non-linear character, turns out to be also
a serious drawback.

Once the TICC2 solution has been properly identified, it is time to
comment on the physical results, going back to Table 1. To this end, we
should have in mind that a nuclear bosonic system with the central S3
interaction tends to collapse when the number of particles $N$ increases,
so that it magnifies the role of the correlations. In Table 1 we can see
that going from TICI2 to TICC2 results in an increasing gain of energy,
ranging from 0.18 MeV for $N=4$ to 963.24 MeV for $N=40$. The importance
of the non-linear terms contained in TICC2 increases with the number of
particles.

It is also interesting to compare with the results provided by an hybrid
ansatz which includes pair correlations as well as hyperradial excitations
in a TICI2 scheme \cite{gua96,gua966}. The combination of two-particle
plus hyperradial correlations is the base of the hyperspherical harmonics
description largely used in few-body systems \cite{fabre,rosati}. The
entry TICI2+HR in Table 1 shows the results obtained by minimizing the
Hamiltonian with respect to the correlation function $f(\rho,r)$, where
$\rho$ is the hyperradius. Such an ansatz is an obvious generalization of
the TICI2 one. The comparison shows that the gain in energy due to
hyperradial excitations is relatively small. TICI2 results change in less
than 2\% when hyperradial correlations are included. The gain due to TICC2
is much more important.

Finally, to complete this analysis it is important to compare our results
with those obtained using a Jastrow correlation function. The entry J--VMC
in Table 1 shows the ground-state energies determined by a variational
Monte Carlo procedure. For $N=16$ and $N=40$ a rough determination of the
ground-state energy was made \cite{gua96} using a simple Jastrow
correlation function $f_J(r) = 1 + a \exp(-br^2)$. For $N=4$ we quote the
Jastrow result given in Ref.\ \cite{bis92}. We can see that our TICC2
results are still much less bound than the Jastrow variational lower
bound. This reveals the importance of three-body and higher order
correlations.

\section{Final comments}

In this paper we have presented a way of determining the ground-state
energy and wave function of a system of bosons described in terms of the
translationally invariant formulation of the coupled-cluster method. The
analysis has been limited to the case of two-body correlations (TICC2
approach), and has been carried out by using Fock space techniques. The
correlation amplitudes satisfy a non-linear coupled set of equations,
which has been explicated here, and has been solved for light and medium
bosonic nuclear systems, considering the S3 semi-realistic interaction of
Afnan and Tang.

To conclude we would like to mention the required further steps and the
pending questions to be solved on line. The first obvious extension of
TICC2 is to consider SUB(3) and presumably SUB(4) truncation, as it is
suggested by the comparison with Jastrow results. Here the main problem to
face up is to find a unique way of describing the translationally
invariant cluster operators, both in coordinate and in Fock spaces. One
may propose that the three-body operators should be a mixture of 1p--1h,
2p--2h and 3p--3h excitations, and the main questions are how to find a
unique description (i.e.\ the object equivalent to ${\cal S}(n)$ of
SUB(2)), and its relation to the characterizing quantum numbers.

In previous studies restricted to light systems (or to heavier nuclei in
the TICI2 approximation) the advantages of the coordinate representation
have been stressed with regard to the actual computations and with respect
to the richness of the correlation operators (i.e.\ the easy way to
introduce spin and isospin dependence when dealing with true nuclei). The
form of the wave function is known [see Eq.\ (\ref{coord})], and it seems
simpler to determine a single function $f(r)$ instead of a large number of
numerical amplitudes ${\cal S}(n)$. The still open problem is to transform
the Schr\"odinger equation with the ansatz (\ref{cc2eq}) into an
appropriate integro-differential equation for the two-body correlation
function.

The coordinate representation of the SUB(2) truncation of CCM is quite
similar to the familiar Jastrow form with only two-body correlations,
where the model state is correlated by the action of the operator
$\prod_{i<j} [1+f(r_{ij})]$. The expansion of this product generates all
terms appearing in Eq.\ (\ref{coord}) and also additional terms which go
beyond the independent-pair character of that equation. It is worth noting
that a form like Eq.\ (\ref{coord}) was used time ago \cite{owen} to
obtain hypernetted-chain equations for the case of state-dependent
correlations. Moreover, the main advantage of the Jastrow ansatz with
respect to the SUB(2) truncation, particularly in the case of strong
interactions, is its ability to tame the potential for all nucleon pairs.
It has been suggested \cite{cmt10} to use this formal analogy to generate
other truncation schemes for CCM which incorporate a Jastrow-like
behaviour, and specially adapted to strongly short-distance repulsive
interactions.

The last open question is to find a TICC2 formulation for fermionic
systems. Here again we expect serious algebraic difficulties related to
the structure of the reference state, with several occupied shells and the
corresponding complexity of the $S^{(1,2)}$ operator. In our opinion, the
formulation in coordinate space will simplify the problems related to the
fermionic statistics.

\vspace{1truecm}

{\bf Acknowledgements}

This work was partially supported by DGICyT (Spain) grant PB92-0820. I.M.\
and M.P.\ acknowledge the DGICyT (Spain) and ICI (Spain), respectively,
for a fellowship.

\newpage

\begin{table}
\caption{
Binding energies (in MeV) of $N$ bosons using the S3 interaction.}
\vspace{0.4cm}
\begin{center}
\begin{tabular}{lccc}
\hline
& $N=4$ & $N=16$ & $N=40$ \\
\hline 
TICC2 & $-$25.49 & $-$1234.86 & $-$8456.55 \\
TICI2 & $-$25.31 & $-$1130.94 & $-$7493.31 \\
TICI2 + HR & $-$25.99 & $-$1145.11 & $-$7555.12 \\
J-VMC & $-$27.15 & $-$1403$\pm$1.5 & $-$9570$\pm$7 \\
\hline
\end{tabular}
\end{center}
\end{table}

\vspace{4truecm}

\section*{Figure captions}

{\bf Fig.\ 1.} 
Energies (in MeV) of two selected TICC2 solutions (solid and long-dashed
lines), plotted as a function of the number $N_{\rm max}$ of equations
considered in the system (\ref{cc2eq}). The TICI2 solution is also
displayed as the short-dashed line.

\noindent {\bf Fig.\ 2.} 
Correlation functions $f(r)$ as defined in Eq.\ (\ref{laefe}) for the
solutions of Fig.\ 1.

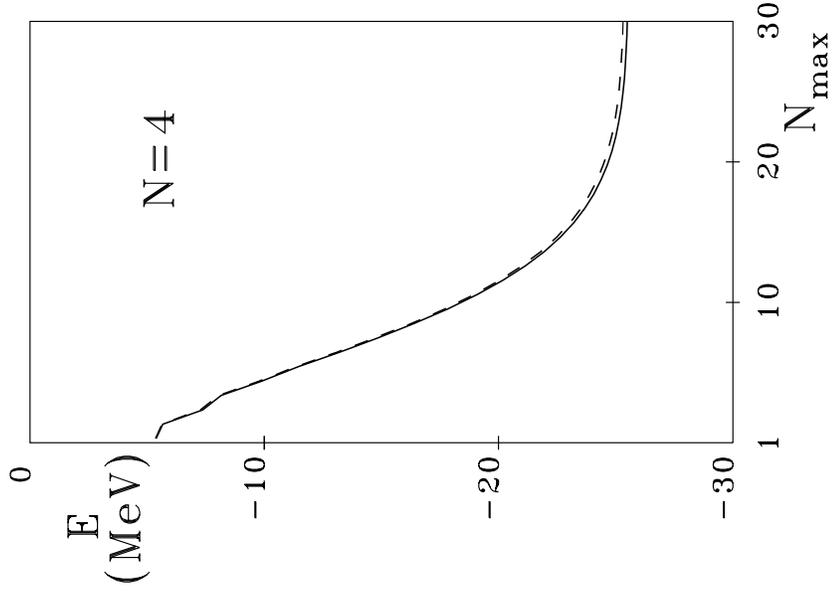
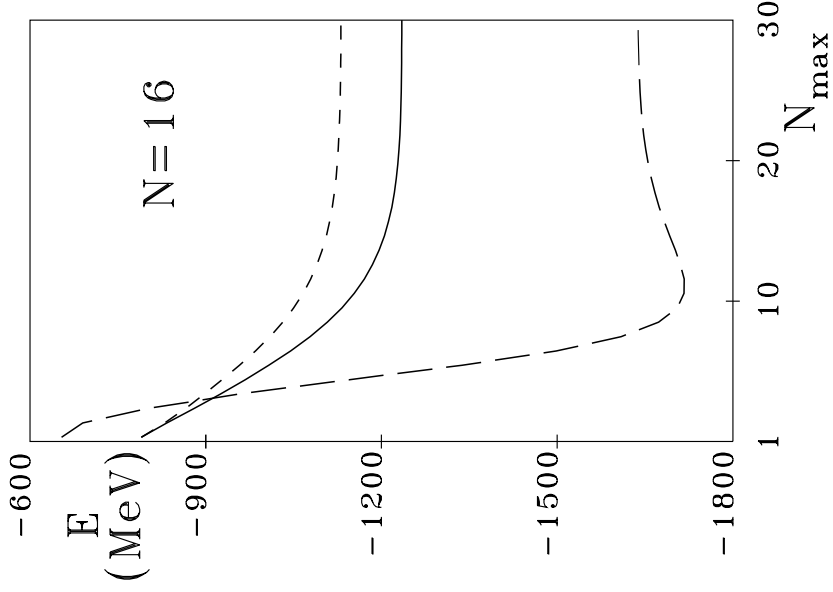
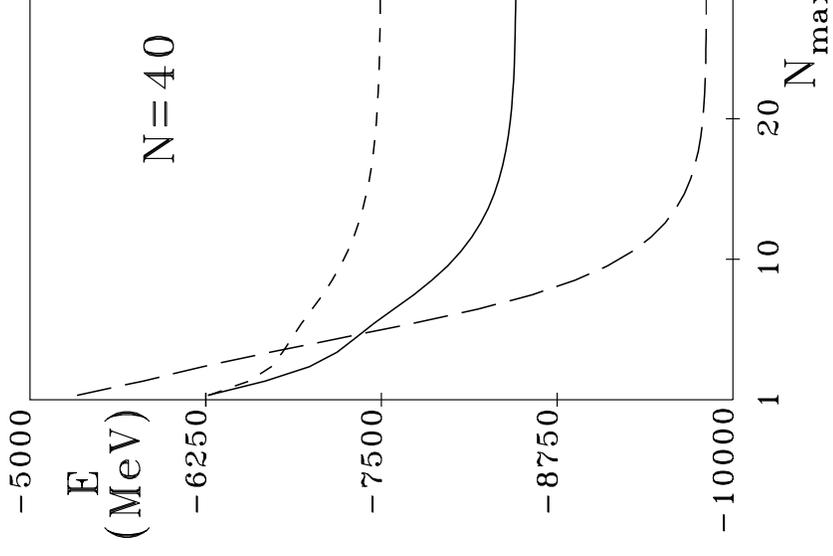
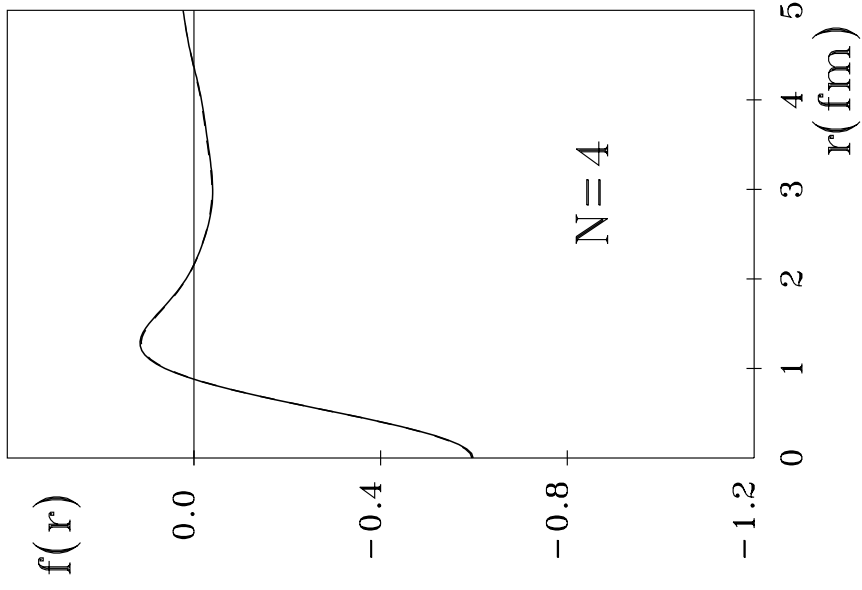
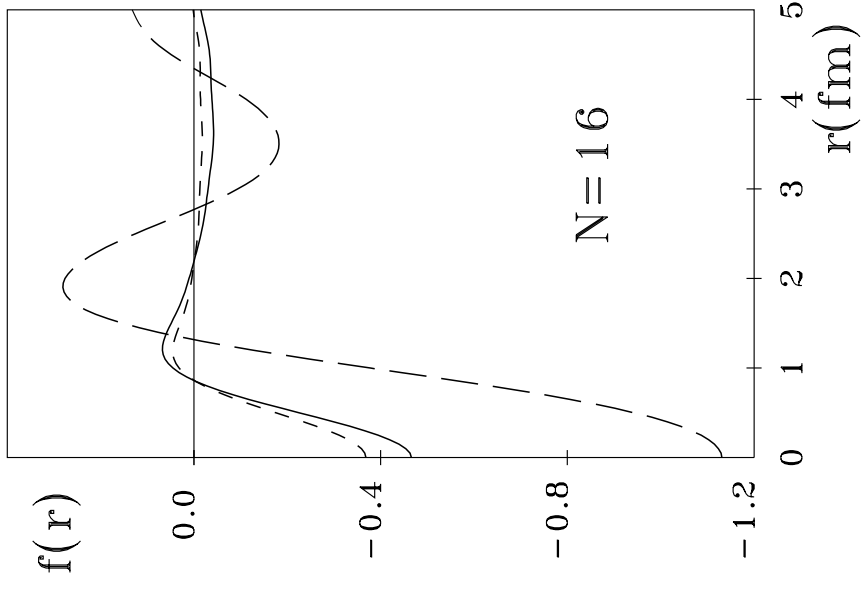
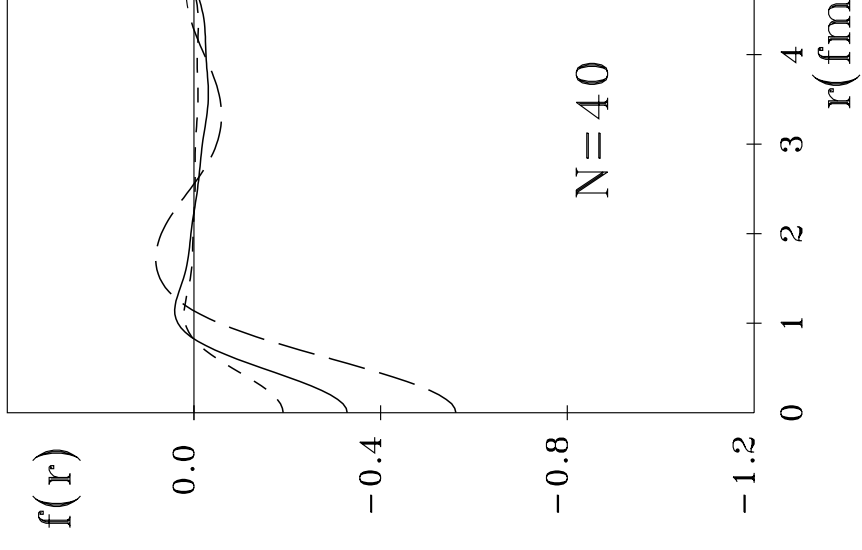
\end{document}